# Review of Considerations for Mobile Device based Secure Access to Financial Services and Risk Handling Strategy for CIOs, CISOs and CTOs

**Amal Saha**[*]
Tata Institute of Fundamental Research (TIFR), Mumbai, INDIA,
Email: amal.k.saha@gmail.com

**Sugata Sanyal**
Tata Consultancy Services (TCS), Mumbai, INDIA
Email: sugata.sanyal@tcs.com

*Corresponding Author*

## Abstract

The information technology and security stakeholders like CIOs, CISOs and CTOs in financial services organization are often asked to identify the risks with mobile computing channel for financial services that they support. They are also asked to come up with approaches for handling risks, define risk acceptance level and mitigate them. This requires them to articulate strategy for supporting a huge variety of mobile devices from various vendors with different operating systems and hardware platforms and at the same time stay within the accepted risk level. These articulations should be captured in information security policy document or other suitable document of financial services organization like banks, payment service provider, etc. While risks and mitigation approaches are available from multiple sources, the senior stakeholders may find it challenging to articulate the issues in a comprehensive manner for sharing with business owners and other technology stakeholders. This paper reviews the current research that addresses the issues mentioned above and articulates a strategy that the senior stakeholders may use in their organization. It is assumed that this type of comprehensive strategy guide for senior stakeholders is not readily available and CIOs, CISOs and CTOs would find this paper to be very useful.

## 1. Introduction

Mobile banking and payments are new convenient schemes for customers to perform transactions, and are predicted to increase rapidly as the number of mobile phone users increases. The use of mobile devices, such as cellular phones, tablets, laptops and personal digital assistants to make payments and access financial services is becoming very common.

However there are risks associated with the usage and they must be taken into consideration. This paper reviews the threat model, risk assessment and discusses practical strategies which financial service providing organizations may consider for selection of supportable mobile devices and security constraints to be imposed on them. Mobile device based financial service providers face these questions and have to address them while formulating enterprise architecture, particularly the digitization of the processes.

Financial services support multi-channel architecture for access and mobile based access channel is becoming a very important one. And security is of paramount importance in financial services like banking, payment and the like. Chief Information Officers (CIOs), Chief Technology Officers (CTOs), Chief Information Security Officers (CISOs) and Information Security Architects at financial services organizations are often asked to articulate important considerations for mobile device based secure access to financial services they provide and strategy for selection of supporting mobile devices. This paper is expected to guide these senior stakeholders to make the right strategic decisions. While classification risks, threat modelling and mitigation approaches are available from multiple sources, the senior stakeholders often find it difficult to articulate policy for handling them in a comprehensive manner for sharing with business owners and other technology stakeholders.

## 2. Review of Threat Model of Mobile Device Usage

There are so many hardware and software platforms under control of so many vendors in case of mobile devices and it makes managing security of the solution very difficult for an organization that needs to support a large number of devices, e.g., financial service providers. Furthermore, comprehensive security must take into account that of the application running on the operating system of the mobile device, the web services on the cloud or server, the integration with remote messaging services (e.g., Google Cloud Messaging, Apple iCloud) and the operating system, the hardware and firmware and the wireless network connection. Mobile devices use Bluetooth, NFC, IEEE 802.11 for short-range radio communication in Wireless LAN (WLAN) and Wi-Fi for data connection, mobile network operator provided data connection (GPRS or 2.5G, 3G, 4G, etc) and context-switched connection for voice and SMS and also USB port based physical connection and they too introduce vulnerabilities.

There are different classifications of threats and vulnerabilities associated with mobile devices. NIST (SP 800-124 Rev 1) [1] categorized the vulnerabilities into the following buckets:

- lack of physical security controls  - loss/theft of device and data
- use of untrusted mobile devices - lack of root-of-trust features and jail-breaking and rooting which bypass built-in security controls
- use of untrusted networks - public communications are susceptible to eavesdropping and man-in-the-middle attacks, use of insecure configuration of protocol (e.g., weaker authentication and encryption algorithms in Wi-Fi ) and non-usage of VPN and private network
- use of untrusted applications - using third-party insufficiently vetted applications from application stores
- interaction with other systems - mobile devices may interact with other systems for data exchange (synchronization) and storage, resulting in data loss and injection of malware

- use of untrusted content - automated processing of URL represented by QR code and obfuscated URL, without user intervention leading to using malicious website, non-deployment of secure web gateway

- use of location service - location services enabled on mobile devices could increase risk of targeted attacks

OWASP [2] identified a set of risks and associated vulnerabilities resulting in the risk:

- spoofing - improper session handling, social engineering, malicious QR code, untrusted NFC tag or peer, malicious application, weak authentication, weak authorization
- tampering - modifying local data, carrier network breach, insecure Wi-Fi network
- repudiation - missing device, toll fraud, malware, client-side injection
- information disclosure - malware, lost device, app reverse engineering, backend service breach
- elevation of privilege - rooted/jail-broken device, compromised device with rootkits, sandbox escape, compromised credentials, flawed authentication, weak authorization
- denial of service - push notification flooding, crashing applications, excessive usage of API

A closer analysis will reveal that both classifications have many things in common and they may be used together. It may be highlighted that the risks are contributed by hardware, firmware, operating system, applications and also network.

## 3. Risks Handling Approaches for Mobile Device Usage in Financial Transactions

Vulnerabilities and associated threats have some probability of occurrence and cause an impact on business. The latter is known as risk to business folks. Risk may be addressed in four different ways: A solution may accept (do nothing), avoid (do not have process or operation that introduces risk), mitigate (put controls), or transfer (outsource or insure) risk.

**Mitigate risk** – activities with a high likelihood of occurring, but financial impact is small. The best response is to use management control systems to reduce the risk of potential loss.
**Avoid risk** – activities with a high likelihood of loss and large financial impact. The best response is to avoid the activity. This is to say that we do not want to engage in such activity or support such associated features in our products or solutions.
**Transfer risk** – activities with low probability of occurring, but with a large financial impact. The best response is to transfer a portion or all of the risk to a third party by purchasing insurance, hedging, outsourcing, or entering into partnerships.
**Accept risk** – if cost-benefit analysis determines the cost to mitigate risk is higher than cost to bear the risk, then the best response is to accept and continually monitor the risk.

For financial transactions which leverage mobile devices, mitigation is often the preferred strategy for handling the risk, with acceptance of residual risks. Henceforth, let us focus of discussion on feasible approaches for mitigation of important risks identified by the stakeholders.

Not articulating the risk acceptance often leads to issue in relationship between technology stakeholders and business owners.

In the following sections, let us discuss capability of mitigating some vulnerabilities which were identified in previous section, at the device hardware and firmware, operating system and application levels. This will help one determine suitability of a device for performing financial transaction safely.

### 3.1 Untrusted Mobile Devices and Building Trust at Device Level

Root-of-trust (RoT) [3] is a collection of fundamental security primitives and capabilities needed to make mobile devices more secure and is often employed in workplace, i.e., enterprise environment, to support Bring Your Own Device (BYOD) scheme. However, the fundamental features of root-of-trust may be leveraged in a mobile device even when it is not used in enterprise environment. For example, financial service providers like banks may also leverage this to build separate trusted environment, with the help of RoT [29].

Fundamentally speaking, when a device is powered on, a hardware or firmware based root-of-trust component measures BIOS of the device from integrity perspective. If the measurements meet expectations, BIOS is executed and booting starts. If the measurements do not match expectations, impacted bad module is rolled back to last known good copy and only then BIOS is executed and booting starts.

Root-of-trust has been implemented in Apple iPhone, Windows Mobile and Samsung's Android smartphones with Knox components. In fact, Blackberry smartphones which once enjoyed high adoption in enterprise space because of higher security standards, lost to competition from iPhones, Windows Mobile and Samsung Knox based smartphones, once they built root-of-trust and other security capabilities in the devices that they manufactured.

Root-of-trust along with secure booting goes a long way in preventing malware injection. It may be pointed out that one of the biggest risks in mobile financial computing is interception of user inputs by malware and sensitive information like account numbers, passwords, PINs, etc may be stolen by malware and transmitted to the command-and-control center operated by criminals.

Many mobile devices, particularly those that are personally owned, are not necessarily trustworthy. Most currently distributed mobile devices lack the root-of-trust features (e.g., trusted platform module or TPM) that are increasingly being built into laptops and other types of hosts. There is also frequent jail-breaking and rooting of mobile devices, which means that the built-in restrictions on operating system have been bypassed and this renders the device less secure.

However, penetration of mobile devices with RoT feature is low and may take a few years to pick up.

### 3.2 Risk Mitigation Approaches and Comparison of Devices in the Market

#### 3.2.1 Sandboxing

Payment solution provider should assume that all phones are untrusted unless it has properly secured them before user access and monitors them continuously. There is a technical solution approach for achieving degrees of trust without using root-of-trust component, such as running the financial applications in a secure, isolated sandbox or container on the mobile device.

A sandbox (or container) is a security mechanism for separating running programs on mobile device and it relies on isolating code and the impact that code can have on runtime environment of mobile device. An abstraction layer is thus provided so that one application cannot step on toes of another application or corrupt the operating system. Typically it goes beyond simple application abstraction and includes encryption for data-at-rest and data-in-motion, with enhanced policy control that goes above and beyond the standard application level control. Secure containers separate financial application data and processing from personal data on the mobile device and prevent critical data from leaking out to unauthorized individuals. This is done by encrypting the data on the mobile device and providing additional data security features, such as copy-paste data loss prevention. A secure container often enables companies to perform a "remote-wipe" of enterprise data controlled by the sandbox.

Sandboxing is not governed by a widely accepted specification and may be classified in the following categories [14, 18].

Bare metal sandboxing requires a hypervisor that typically lives at the firmware level. Above this firmware-based hypervisor, there is one or more virtualized operating systems which are completely sandboxed from one another. This is the most secure approach, but requires significant work by device manufacturers and the technology did not see any significant traction in consumer smartphone and tablet space.

Second category of sandboxing runs on top of an existing operating system such as Windows Mobile or Android. Device manufacturers need not do significant work to support this type of sandboxing and it usually requires some small change in the operating system and no modification at the firmware level is needed. The downside is consumption of more resources by the "host" operating system, which runs one or more virtual (or guest) operating systems on top. Even for this one too, industry did not see significant adoption.

The third category of sandboxing is concerned with application level isolation where a standard application, which uses the native OS environment and API's, is used to provide a sandboxed environment for other selected applications and data, and introduces additional security controls like advanced policy control, encryption, digital signature, etc. This category of sandboxing is currently very important for the mobile industry since there are no standards for the other two. This category is perhaps the best way to provide sandboxing, given the current state of adoption of mobile device technologies. Adoption of other two categories of sandboxing would probably take a few years.

Application level sandboxing can be broken down into the following sub-categories - content wrapper, workspace wrapper and application wrapper.

Content wrapper is focused on providing a secure container for enterprise documents. Many variants of content wrappers are becoming popular and are often directly aimed at stemming the document-sharing problem.

Workspace wrapper is aimed at providing a full work space environment for enterprises, including email, calendar, secure browsing, document editing, etc. Some workspace wrapping sandboxes are now allowing organizations to embed their own home-grown and enterprise applications inside the sandbox, ensuring that there is one common entry point to access any enterprise application or data.

Application wrapper introduces additional security layer, usually in the form of a software development kit or API that developers can integrate with their application and not have to worry about implementing their own data-at-rest and data-in-motion encryption. Typically this technology provides additional policy control mechanisms which allow wrapped application to be individually wiped out thru enterprise Mobile Device Management system without touching the rest of the mobile device. From the perspective of information disclosure by malware in financial services applications, application wrapper would be most relevant.

A container software is installed remotely by a trusted 3rd-party on such mobile device as a one-time activity, with concurrence of the user.

The service administration team working for service providers can set a policy to encrypt data outside and inside the sandbox or container.

The sandboxing, among other things, typically relies on authentication and authorization (access control) of the mobile device by network access control (NAC) layer [15]. When a mobile device connects to a computer network, it is not permitted to access anything unless it complies with a business defined policy including anti-virus protection level, system update level and configuration. While the computer or the mobile device is being checked by a pre-installed software agent, it can only access resources that can remediate (resolve or update) any issues. Once the policy is met, the computer is able to access network resources, within the policies defined within the NAC system. NAC is mainly used for endpoint health checks and also Role Based Access. Access to the network will be given according to profile of the person or the mobile device and the results of a posture/health check. NAC solutions allow network operators to define policies, such as the types of mobile devices or roles of users allowed to access areas of the network, and enforce them in switches, routers, firewalls, etc. Where conventional IP networks enforce access policies in terms of IP addresses, NAC environments attempt to do so based on authenticated user or device identities, at least for user end-stations such as laptops and desktop computers.

Combination of Mobile Device Management and NAC techniques would help secure the system by enforcing identity validation, access control and preventing information disclosure.

Samsung Knox [16] enabled mobile devices, Blackberry10, Windows Mobile smartphones, iPhones, etc leverage sandboxing and root-of-trust. Most of these devices are FIPS 140-2 [8] certified (often at level 1). They have secure booting, integrity checks and trusted execution environment of some kind and they can create a separate secure container or sandbox.

### 3.2.2 Connected Small Form-Factor Physical Token

The token be used not just user authentication, but also for transaction integrity and non-repudiation (through PKI) on untrusted mobile devices. This however impacts user experience because mandating carrying a physical token is usually considered unfriendly.

Connected tokens are tokens that must be physically connected to the computer or mobile device with which the user is authenticating. Tokens in this category automatically transmit the authentication information to the client computer once a physical connection is made, eliminating the need for the user to manually enter the authentication information. However, in order to use a connected token, the appropriate input device must be installed. The most common types of physical tokens are smart cards and USB tokens, which require a smart card reader and a USB port respectively.

Smart cards are designed to protect the information they contain [4]. Tamper resistance techniques are used to protect the contents of the chip embedded on the card. Because standard credit card-size smart cards require a reader can somehow interface with the mobile device, it is not useful for most mobile devices. These days many mobile devices are supporting USB port and hence USB Smartcard is a practical portable and multi-platform strong user authentication solution for most mobile devices not equipped with root-of-trust.

Smart-card-based USB tokens which contain a smart card chip inside provide the functionality of both USB tokens and smart cards. Besides authentication of mobile device, authentication, i.e., validation of identity, of the user of the device is needed.

### 3.2.3 Soft Token and SMS based OTP

SMS and soft token based OTP solutions [5, 6] are not purely out-of-band because they still use context-switched 2G/3G/4G connection and data connection, respectively, on the mobile device and this can be intercepted by malware. Hence this approach is less secure than hardware token based solution which does not use any of the connections mentioned above and hence cannot be intercepted by malware. Soft token and SMS based OTPs are often used because of convenience.

### 3.2.4 Device Fingerprinting

A user device is associated with a dynamic trust score [10, 11, 12, 13] that is calculated based on various activities and information associated with the mobile device including the configurations. The computation could use parameters of the device, such as device type, registered device location, the last time the device has been accessed, device phone number, device ID, etc and activities the device engages in, such as value of transactions, value of denied requests, value of approved requests, location of requests, etc. Based on a transaction request from the user device, the trust score and a network reputation score may be used to determine an overall trust score associated with the transaction request.

Device fingerprinting can help uniquely identify the device and distinguish malicious and non-malicious ones. Mobile device attributes like operating system and browser type, etc may have subtle differences among used devices, e.g., browser plug-ins and other attributes which may be less obvious.

Device fingerprinting approach may detect malicious devices regardless of credit card, name or IP address used. It is valuable because fraudsters use stolen identities and proxies to bypass IP address blacklists and IP geolocation filtering mechanisms employed by many financial service providers or their services partners. As a fraud prevention tool, real value of device fingerprinting lies in the ability to transparently correlate device attributes and anomalies at the browser, packet, protocol and OS levels in order to detect fraud attempts. This can work even if the mobile device is not equipped with root-of-trust (RoT) security features and most devices in the market are not having RoT.

Financial application developers may integrate with third-party device fingerprinting solution and SDK on client side programming and employs integration with solution provider accumulated SaaS based device score related services on the server side.

### 3.2.5 Geo-location Check

Checks may be introduced to stop a transaction originating from mobile device in a specific geo-location. Solutions available in the industry typically combine geo-location and device fingerprinting checks.

### 3.2.6 Data-in-motion Protection

VPN for connection between the mobile device and the enterprise is a commonly deployed mechanism for data protection at channel level.

When VPN cannot be used to cover authentication and encryption for the session across applications, SSL/TLS for communication channel with Diffie-Hellman key exchange [17] may be considered at application level. This mechanism may be used in application to exchange data over a secure channel.

Message level encryption may be applied to ensure further protection of data-in-motion.
Derived unique key per transaction (DUKPT) key management [7] with FIPS 140-2 Level 3 [8] certified crypto-module, e.g., Secure Element, may be used on client side, for protection of data-in-motion. DUKPT is a key management technique and is commonly used for POS terminals [9], typically for 3DES encryption algorithm. Most mobile devices do not support software-only DUKPT because it is not secure and requires a capable hardware component to be inserted to the mobile device thru USB port or audio jack.

For laptop TPM (trusted platform module) may be leveraged for such scheme. For mobile phone and tablet, dongle may be attached from outside or Secure Element with adequately hardened and certified (SIM, micro-SD or embedded) may be used.

### 3.2.7 Application Level Intrusion Detection and Web Application Firewall for Mobile Security

Most mobile interactions with the server use HTTP and HTTPS protocols and the applications may leverage intrinsic security controls built into the web applications, web application firewall (WAF) and application level intrusion detection and prevention techniques [27]. Miscellaneous other controls have been employed [24, 25, 26]. Use of WAF and intrinsic security controls is very common and PCI DSS and PA-DSS [28] and other compliances also enforce deployment of such controls.

## 4. Mobile Device Selection - Choices before CIOs, CTOs and CISOs in Financial Service Organizations

Can all mobile devices be supported for delivery of financial services in a secure way? Should the high-end devices only be supported, thereby excluding a large number of mobile device owners? Which security controls, out of those reviewed in the paper, are considered absolutely necessary by the financial services organization's information management team? Which mobile devices, along with third-party solutions in deployment architecture, are capable of supporting those security controls? Which risks may be accepted? Which risks should be mitigated? Should one consider different transaction limit for devices with different degree of security controls? How would one identify device security posture? Should financial services organization invest in mobile device management (MDM) [19] and/or secure web gateways (SWGs) [20] solutions for protecting customers (not employees)? These are some of the common questions that need to be addressed by the office of information officer and security architects. Often financial services organization may not have a well-defined strategy to address them.

MDMs and SWGs are typically employed by enterprise to manage devices used by the employees and not for non-enterprise users. Hence these solutions are typically not employed by financial services firms to secure mobile financial transaction by customers. Often CIOs are not sure if they should adopt these solutions.

Device fingerprinting, geo-location checks and related security features are often by third-party hosted solution providers [21, 22, 23]. Since trusted mobile computing with RoT features, is not going to get wide acceptance in next 4-5 years, device fingerprint and associated checks are likely get wide acceptance for supporting wide range of mobile devices in financial services industry. Therefore device fingerprinting is going to stay as a dominant anti-fraud measure in mobile based access to financial services in future.

Mobile platforms iOS, RIM/Blackberry, Java, Android, Windows Mobile and access channels like SMS, USSD, Mobile Web and Rich Client may have to be supported by the financial services organization. All channels and devices will not have similar security posture. For example, encryption of data between the mobile device and the BTS (tower) in GSM network thru USSD may be cracked easily and transaction limit for this cannel cannot be high. Most devices are not equipped with RoT and we may use device fingerprinting instead. But the latter may not be cheap to implement. Here defining risk acceptance level in organization information security policy and getting approval from business owner, would be important.

Formal risk handling and threat modelling for supporting various mobile devices in financial services industry may be developed by the office of CIOs, CISOs and Security Architects and the strategic approaches highlighted by the authors may be leveraged. They may capture them in information security policy document or other artifacts.

## 5. Conclusions

Mobile platforms iOS, RIM/Blackberry, Java, Android, Windows Mobile and access modes SMS, Mobile Web and Rich Client may be supported within defined risk acceptance level. And security parameter like transaction limit, authentication level, authorization level and confidentiality of the operation may be set appropriately, thru a formal, business-owned risk management practice.

This paper shows approaches for handling risks with mobile computing channel for financial services that are supported. It also shows approaches for defining risk acceptance level and mitigating them. And thereby it helps to articulate strategy for supporting a huge variety of mobile devices from various vendors with different operating systems and hardware platforms and at the same time stay within the accepted risk level. Often these articulations are captured in information security policy document of financial services organization, but many organizations do not have a formalized approach. Though risks and mitigation approaches are often available from multiple sources, senior technology stakeholders would often find it challenging to articulate the issues in a comprehensive manner to business owners of the organization and other technology stakeholders. This paper reviews the current research that addresses the issues mentioned above and articulates a strategy that the senior stakeholders may leverage in their organization. It is assumed that this type of comprehensive strategy guide for senior stakeholders is not readily available and CIOs, CISOs and CTOs would find this paper to be very useful.